\documentclass{article}

\usepackage{PRIMEarxiv}

\usepackage[utf8]{inputenc}
\usepackage[T1]{fontenc}
\usepackage{hyperref}
\usepackage{url}
\usepackage{booktabs}
\usepackage{amsfonts}
\usepackage{amsmath}
\usepackage{nicefrac}
\usepackage{microtype}
\usepackage{fancyhdr}
\usepackage{graphicx}
\usepackage{orcidlink}
\graphicspath{{media/}{figures/}}

\title{I Can't Patch My OT Systems!\\
A Look at CISA's KEVC Workarounds \& Mitigations for OT}

\author{
  Philip Huff\,\orcidlink{0000-0003-0869-2147}, 
  Nishka Gandu\,\orcidlink{0009-0006-1933-9209} \\
  Cyberspace Operations Research and Education Center \\
  University of Arkansas at Little Rock, Little Rock, AR, USA \\
  \texttt{\{pdhuff, ngandu\}@ualr.edu}
  \And
  Pavel Novák\,\orcidlink{0009-0000-5488-2190} \\
  Faculty of Informatics \\
  Masaryk University \\
  Brno, Czech Republic \\
  \texttt{novakpav@mail.muni.cz}
}
\date{v1: Oct 7, 2025}

\begin{document}
\maketitle

\begin{abstract}
We examine the state of publicly available information about known exploitable vulnerabilities applicable to operational technology (OT) environments. Specifically, we analyze the Known Exploitable Vulnerabilities Catalog (KEVC) maintained by US Department of Homeland Security Cybersecurity and Infrastructure Security Agency (CISA) to assess whether currently available data is sufficient for effective and reliable remediation in OT settings. Our team analyzed all KEVC entries through July~2025 to determine the extent to which OT environments can rely on existing remediation recommendations. We found that although most of the entries in the KEVC could affect OT environments, only 13\% include vendor workarounds or mitigations as alternatives to patching. This paper also examines the feasibility of developing such alternatives based on vulnerability and exploit characteristics, and we present early evidence of success with this approach.
\end{abstract}


\section{Introduction}
This article examines the state of publicly available information about known exploitable vulnerabilities applicable to operational technology (OT) environments, such as industrial control systems and other cyber-physical infrastructure. Specifically, we analyze the Known Exploitable Vulnerability Catalog (KEVC) maintained by DHS CISA to assess whether publicly available data is sufficient for effective and reliable remediation in OT environments. Our team analyzed all 1{,}364 vulnerabilities listed in the KEVC through July 2025 to determine the extent to which OT environments can rely on existing remediation recommendations.

The Known Exploitable Vulnerability Catalog (KEVC) identifies vulnerabilities actively exploited by adversaries, including nation-states and ransomware groups. The KEVC is maintained by the Cybersecurity and Infrastructure Security Agency (CISA) within the U.S. Department of Homeland Security (DHS) and is widely used for prioritizing vulnerability remediation, as the sheer volume of Common Vulnerabilities and Exposures (CVEs) continues to grow. Over 40{,}000 CVEs are publicly disclosed each year, but only about 200 per year are confirmed to be actively exploited.

In OT environments, prioritizing remediation is crucial due to the high operational costs and complexity associated with ensuring system safety, reliability, and productivity. OT systems are often purpose-built to support critical physical processes, and changes to these systems are both disruptive and costly. Extensive planning, downtime, and rigorous regression testing are necessary to maintain operational excellence. Traditional ``patch early, patch often'' guidance is rarely feasible.

At the core of the vulnerability management ecosystem lies the CVE system, which is maintained by a global network of CVE Numbering Authorities (CNAs). Currently, 467 CNAs across 40 countries, primarily software vendors, contribute to this consortium, with operational oversight from MITRE and DHS \cite{CVE_CNAs}. Although CVEs provide identification and description of vulnerabilities, they do not provide direct remediation guidance.

Remediation strategies are typically communicated through vendor-issued advisories or open-source community notifications. These advisories inform consumers about affected software versions and offer remediation solutions, usually through new software releases or specific patches. Despite their importance, no comprehensive global advisory database exists. However, software suppliers may adopt open standards for machine-readable advisories. The most widely used format is the Common Security Advisory Framework (CSAF) maintained by OASIS \cite{oasis_csaf20}, which succeeded the Common Vulnerability Reporting Framework (CVRF). Similar frameworks, such as the Open Source Vulnerabilities (OSV) project managed by the Open Source Security Foundation (OpenSSF), target open-source packages. Additionally, the Vulnerability Exploitability eXchange (VEX) standard complements these by clearly indicating software versions and configurations vulnerable to specific CVEs \cite{cisa_vex_minreqs}.

CSAF is particularly relevant to OT environments because it documents machine-readable remediation strategies, including workarounds and mitigations. According to CSAF definitions, a \emph{workaround} provides guidance on configurations or deployment scenarios that prevent vulnerability exposure, while a \emph{mitigation} reduces vulnerability risk without fully resolving it \cite{oasis_csaf20}. A minimally effective vulnerability management process in OT involves monitoring the KEVC for actively exploited vulnerabilities, cross-referencing these with available vendor advisories, and employing documented workarounds or mitigations to develop practical remediation playbooks. However, several barriers undermine this ideal workflow. First, accurately identifying and correlating OT assets with applicable CVEs remains a challenge, and most organizations struggle to maintain comprehensive inventory of OT assets. Second, our analysis reveals that reliable advisories from software vendors are often scarce, and even fewer provide actionable workarounds or mitigations.

This article examines the current efficacy of vendor advisories mapped to the KEVC, identifies existing data gaps, and proposes alternative approaches to enhance the availability of meaningful remediation strategies for OT systems.

\section{Related Works}
Vulnerability remediation in operational technology (OT) has been treated in the literature and practitioner guidance as a problem that is fundamentally different from IT patching, primarily due to the issues mentioned earlier, notably the inability to patch devices regularly. Several major guidance documents stress that remediation must balance security against uptime, safety and vendor-certified configurations rather than blindly applying the fast patch-and-reboot cycle used in IT. NIST’s Guide to OT security explicitly frames OT remediation within this trade-off and recommends layered, risk-aware controls and change-management processes tailored to OT life cycles~\cite{nist_sp800_82_r3}.

IEC 62443~\cite{iec_62443_series} provides a defense-in-depth architecture (zones/conduits) and prescriptive process controls that are widely referenced when designing remediation and patch-management policies for industrial control systems (ICS). Practitioners typically map remediation actions (patching, configuration changes, compensating controls) into an organization’s IEC-aligned zones, using the standard to justify scheduling, testing and segregation choices that reduce operational risk.

CISA supplies OT-specific asset-inventory, vulnerability-discovery, and mitigation guidance that emphasizes continuous asset discovery, prioritized (risk-based) remediation, and the use of compensating controls where immediate patching is infeasible (e.g., network segmentation, virtual patching, protocol hardening, monitored workarounds). These resources also advocate coordinated disclosure and vendor engagement for legacy or end-of-life devices~\cite{cisa2025_ot_asset_inventory}.

Empirical and practitioner literature documents why OT remediation lags IT and surveys approaches that have been tried in the field. Measurement studies of ICS patching show slow, heterogeneous uptake driven by long maintenance windows, vendor approval requirements and the diversity of vendor/firmware ecosystems; vendor/community white papers and incident-response vendors (e.g., Dragos, Armis) describe practical patterns such as risk-based triage, prioritized backlogs, and “virtual” or compensating mitigations to buy time for safe patch deployment. These works collectively argue that remediation is as much an engineering/systems problem (testing, rollback, staging) as a security triage problem~\cite{10.1145/3143314.3078524}.

Current knowledge offers a wealth of advice regarding vulnerability remediation. However, even though some of the guidance includes the possibility of using workarounds~\cite{cisa2008_patch_management_control_systems}, to the best of our knowledge, there is no work systematically mapping and verifying the ability to automatically obtain and apply these workarounds.

\section{Data \& Methodology}
A team of eight cybersecurity research analysts systematically evaluated all 1{,}391 vulnerabilities listed in the KEVC as of August 1, 2025. The primary objective was to assess the completeness, specificity, and practicality of vendor-provided advisories associated with each vulnerability, with a focus on their applicability and effectiveness within OT systems.

Each researcher independently evaluated every vulnerability and recorded whether the vendor:
\begin{enumerate}
\item Issued an official patch or software update.
\item Supplied remediation instructions in a machine-readable format (e.g., CSAF, VEX, JSON, or other structured data suitable for automation).
\item Documented one or more workarounds or mitigations that could be applied when patching was not feasible.
\end{enumerate}

For vulnerabilities where alternative remediation was referenced or reasonably inferred, the analysts captured four additional qualitative attributes:
\begin{enumerate}
\item \textbf{Specificity of Remediation Guidance:} labeled \emph{Detailed}, \emph{Generic}, or \emph{None} according to the level of empirical evidence provided.
\item \textbf{Effectiveness:} assessed the expected risk reduction of recommended remediations.
\item \textbf{Feasibility of Alternative Remediation Generation:} evaluated the feasibility of independently generating practical remediation strategies based solely on publicly available vulnerability details when vendor recommendations were unavailable or insufficient.
\item \textbf{Product Identification and Advisory Clarity:} examined the format and clarity of affected product information within advisories to determine ease of asset mapping and applicability to OT environments and whether advisories clearly differentiated applicable and non-applicable products to prevent misapplication of remediation actions.
\end{enumerate}

Each affected product was further classified according to the United Nations Standard Products and Services Code (UNSPSC). This classification facilitated systematic identification and analysis of vulnerabilities relevant to OT-specific software and firmware.

\begin{figure}[htbp]
  \centering
  \includegraphics[width=\linewidth]{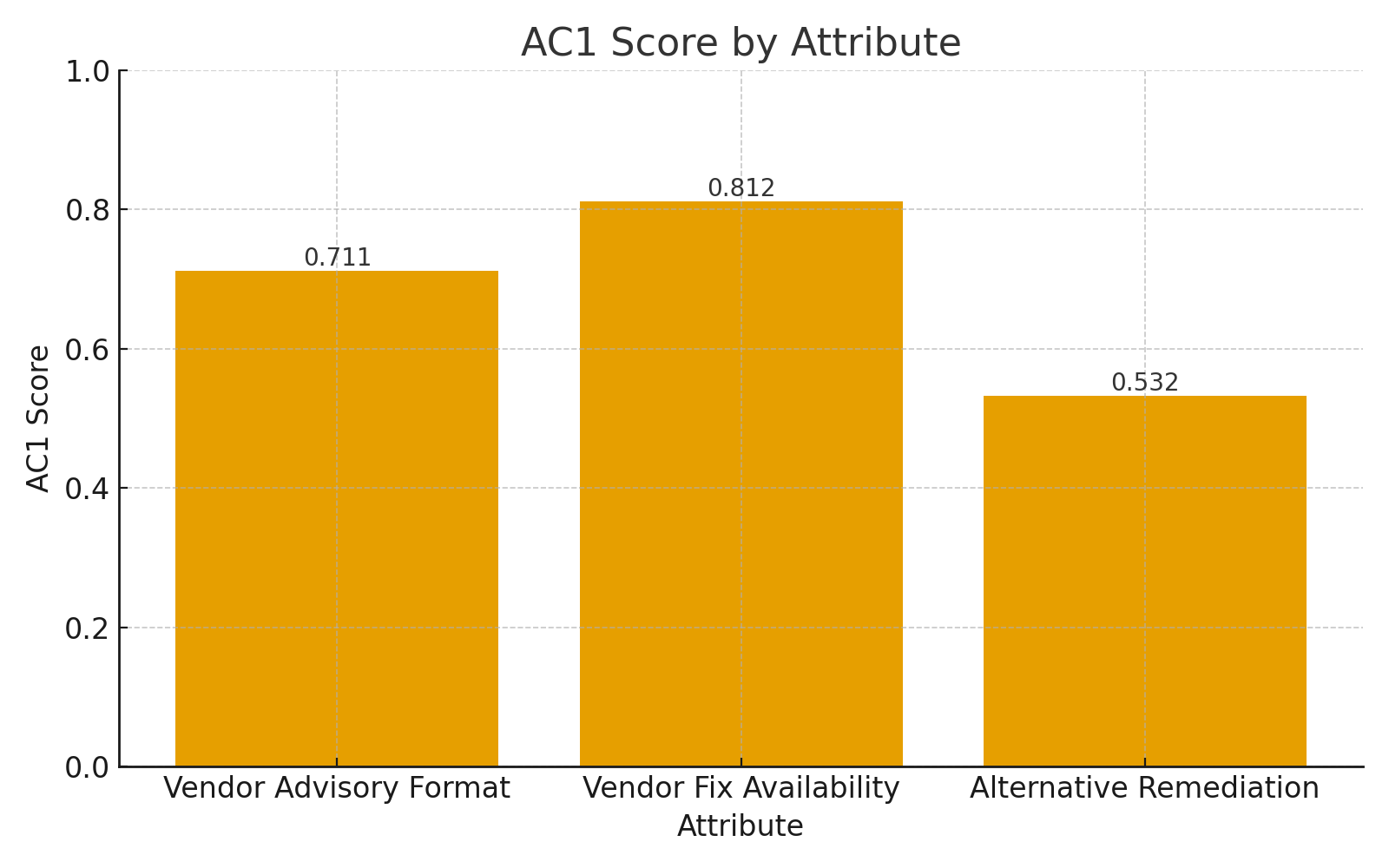}
  \caption{Agreement (AC1) by attribute across an arbitrary 10\% validation sample. Lower scores for vendor advisory format and alternative remediation likely reflect the difficulty of locating advisory information on vendor websites. }
  \label{fig:ac1-score}
\end{figure}

To ensure reliability and validity in our findings, we conducted an inter-rater reliability assessment through an independent validation of a random sample comprising 10\% of the total vulnerabilities evaluated. For inter-rater reliability, we prefer the agreement coefficient for nominal data (Gwet’s AC1), which is well suited for nominal data and avoids the kappa prevalence paradox that arises in data sets dominated by a single characteristic or option. \cite{feinstein1990paradox, gwet2008computing}. Using AC1, we compute:
\begin{equation}
\label{eq:ac1}
AC1 \;=\; \frac{P_o - P_e^{AC1}}{1 - P_e^{AC1}}, \qquad
P_e^{AC1} \;=\; \sum_{i=1}^{k} p_i(1-p_i), \qquad
p_i \;=\; \tfrac{1}{2}\!\left(\frac{n_{i\cdot}}{N} + \frac{n_{\cdot i}}{N}\right),
\end{equation}
where $P_o$ is observed agreement, $N$ is the number of overlapped items, $n_{i\cdot}$ and $n_{\cdot i}$ are the row/column totals for category $i$ from the two raters, and $k$ is the number of categories. The inter-rater assessment showed sufficiently high reliability among the categories shown in Figure~\ref{fig:ac1-score}.

\section{Key Findings}

\subsection{Applicability to OT Environments}
We asked whether software in the KEVC is commonly present in OT environments such as industrial control systems. If KEVC entries rarely apply to OT contexts, the remediation workload specific to OT would be limited. In contrast, broad applicability would imply that sustained effort is needed to address KEVC items in OT programs.

\begin{figure}[htbp]
  \centering
  \includegraphics[width=\linewidth]{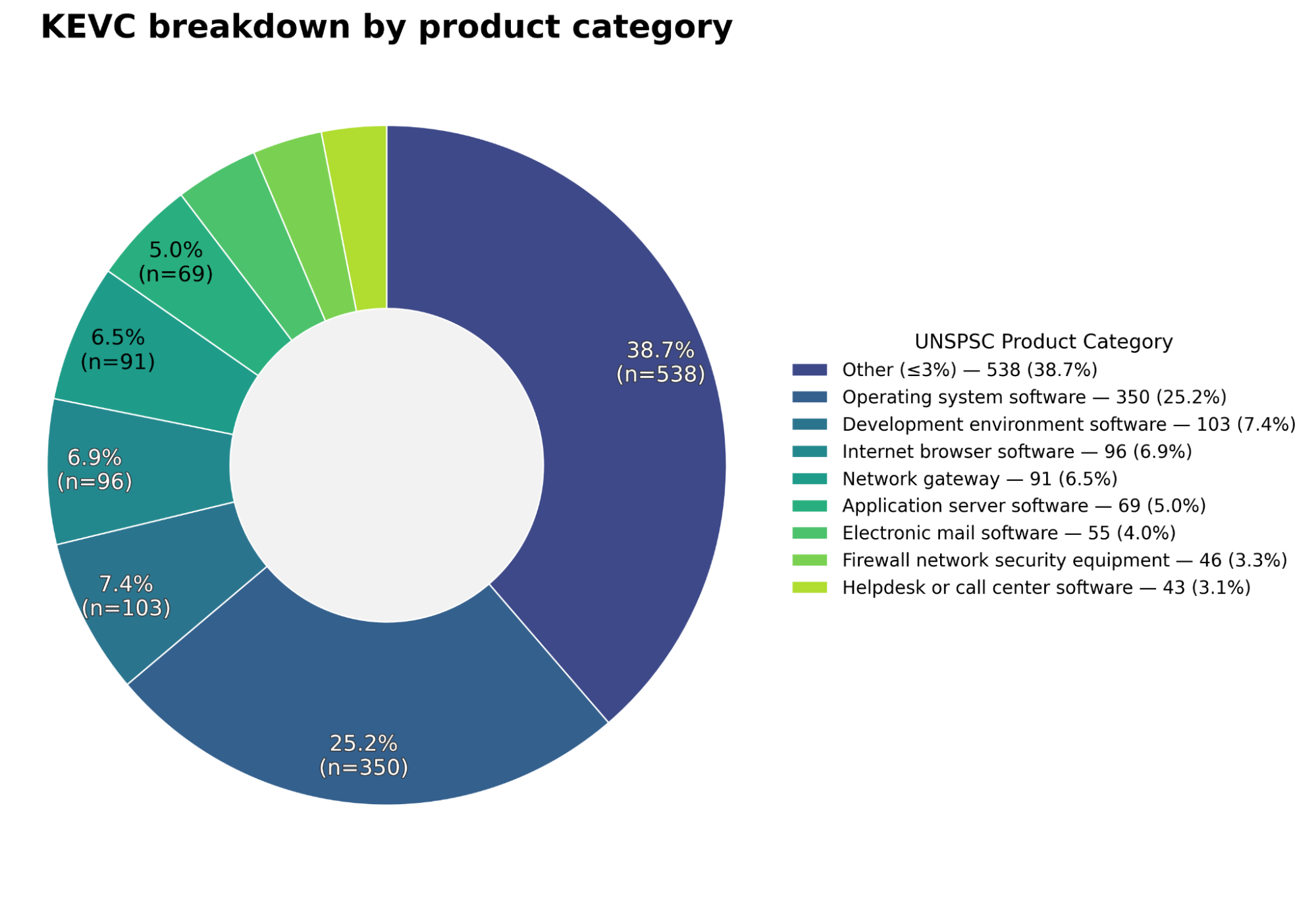}
  \caption{KEVC entries grouped by UNSPSC product categories.}
  \label{fig:kevc-breakdown-by-product}
\end{figure}

To categorize affected software, we mapped each KEVC entry to a United Nations Standard Products and Services Code (UNSPSC) category. We provided the product name and vulnerability description to an OpenAI model \texttt{gpt-4o-2024-08-06} using a standardized prompt that instructed the model to (i) assign the most specific UNSPSC class and category, and (ii) return a brief rationale to assist with validation. We grouped categories representing less than 3\% of entries into ``Other'' to aid readability as shown in Figure~\ref{fig:kevc-breakdown-by-product}.

Our results show \textit{Operating system software} (e.g., core Windows and Linux) accounts for 25.2\% of KEVC entries. Additional categories commonly encountered in OT, such as \textit{Internet Browser Software, Network Gateway, Application Server Software}, and \textit{Firewall Network Security Equipment}, also appeared prominently. Although only two entries were explicitly categorized as \textit{Industrial Control System Software}, many other categories represent software that commonly exists in OT environments to support networking, security, or operator workflows. For example, engineering workstations and other HMI typically include a browser and a PDF viewer. 

\begin{figure}[htb]
  \centering
  \begin{minipage}{0.48\textwidth}
    \centering
    \includegraphics[width=\linewidth]{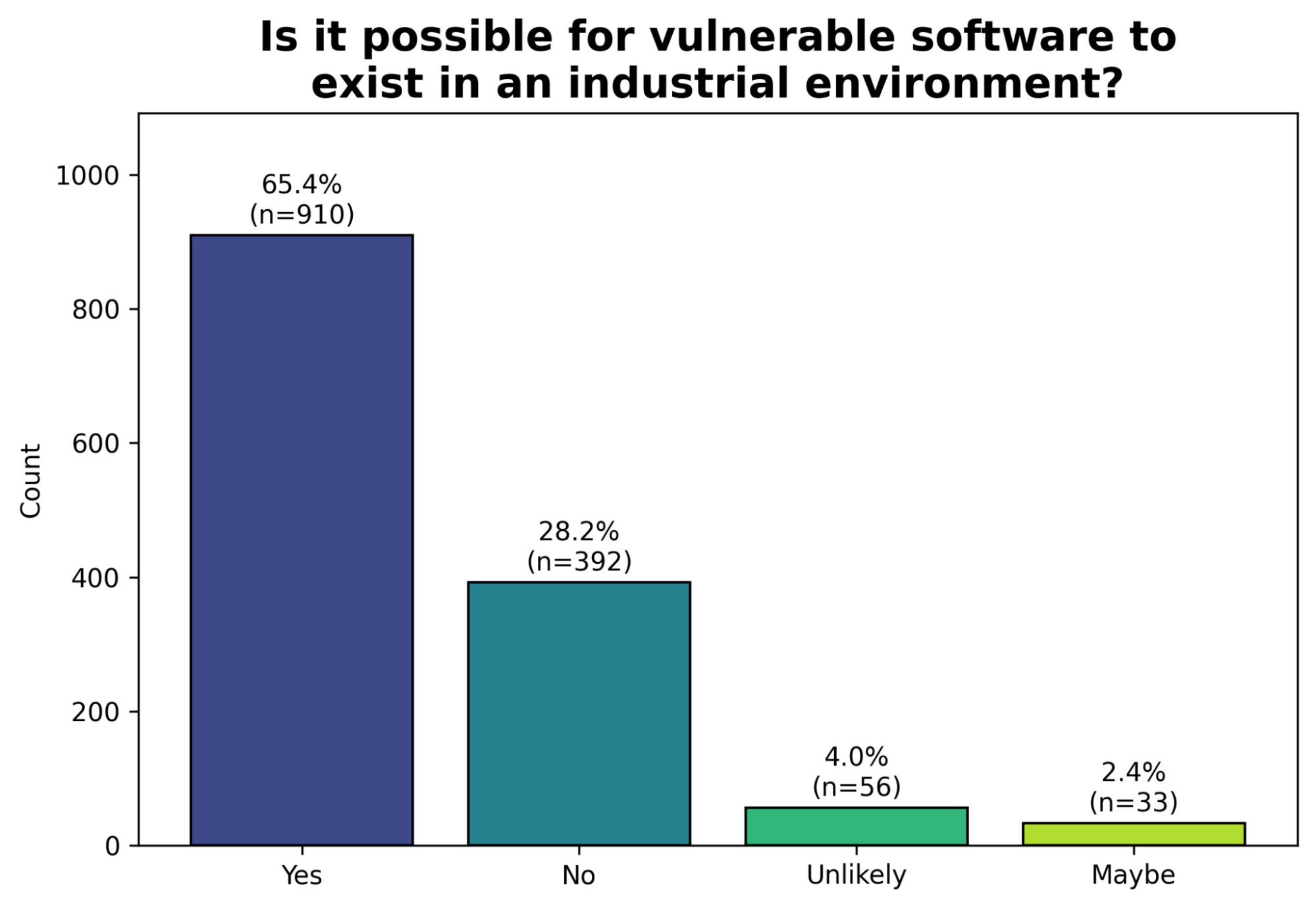}\\
  \end{minipage}\hfill
  \begin{minipage}{0.48\textwidth}
    \centering
    \includegraphics[width=\linewidth]{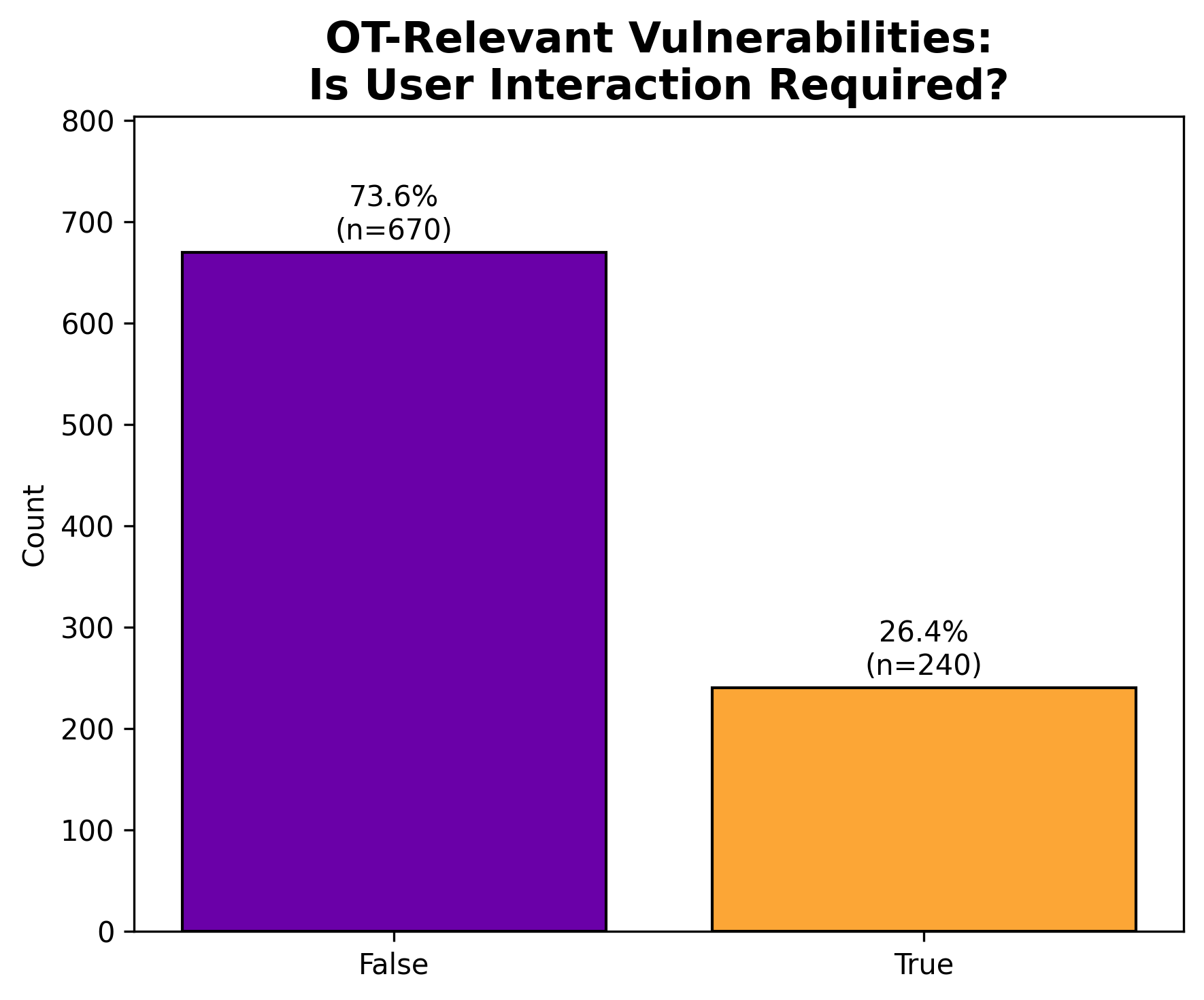}\\
  \end{minipage}
  \caption{Characteristics of KEVC entries in OT environments. 
(Left) Share of entries plausibly present in OT systems. 
(Right) User‐interaction requirements for OT‐relevant entries.}
  \label{fig:kevc-ot}
\end{figure}

Based on the UNSPSC mapping, we estimate that 65.4\% of KEVC entries could plausibly be found in OT environments (Figure~\ref{fig:kevc-ot}, left).  We also examined exploitation conditions for OT-relevant entries. A large majority (73.6\%) do not require user interaction for exploitation (Figure~\ref{fig:kevc-ot}, right). This profile indicates elevated operational risk because many vulnerabilities could be exploited without an operator initiating any action.

These results indicate that exploitable software covered by the KEVC is not only present in OT environments, but that a substantial fraction of the catalog is OT-relevant. Moreover, within that OT-relevant subset, the predominance of non-interactive exploitation pathways suggests elevated operational risk because exploitation does not rely on human interaction and could occur discretely and automatically.

\subsection{Do Vendors Support Patching Alternatives?}
Having shown that KEVC vulnerabilities are present in OT environments, we next assessed how well remediation alternatives are supported by vendors in the KEVC. Because routine patching is often infeasible in OT due to the high cost of downtime and the need for regression testing to protect safety and reliability, operators frequently rely on workarounds and mitigations until a safe patch window becomes available.

\begin{figure}[htbp]
  \centering
  \includegraphics[width=.85\linewidth]{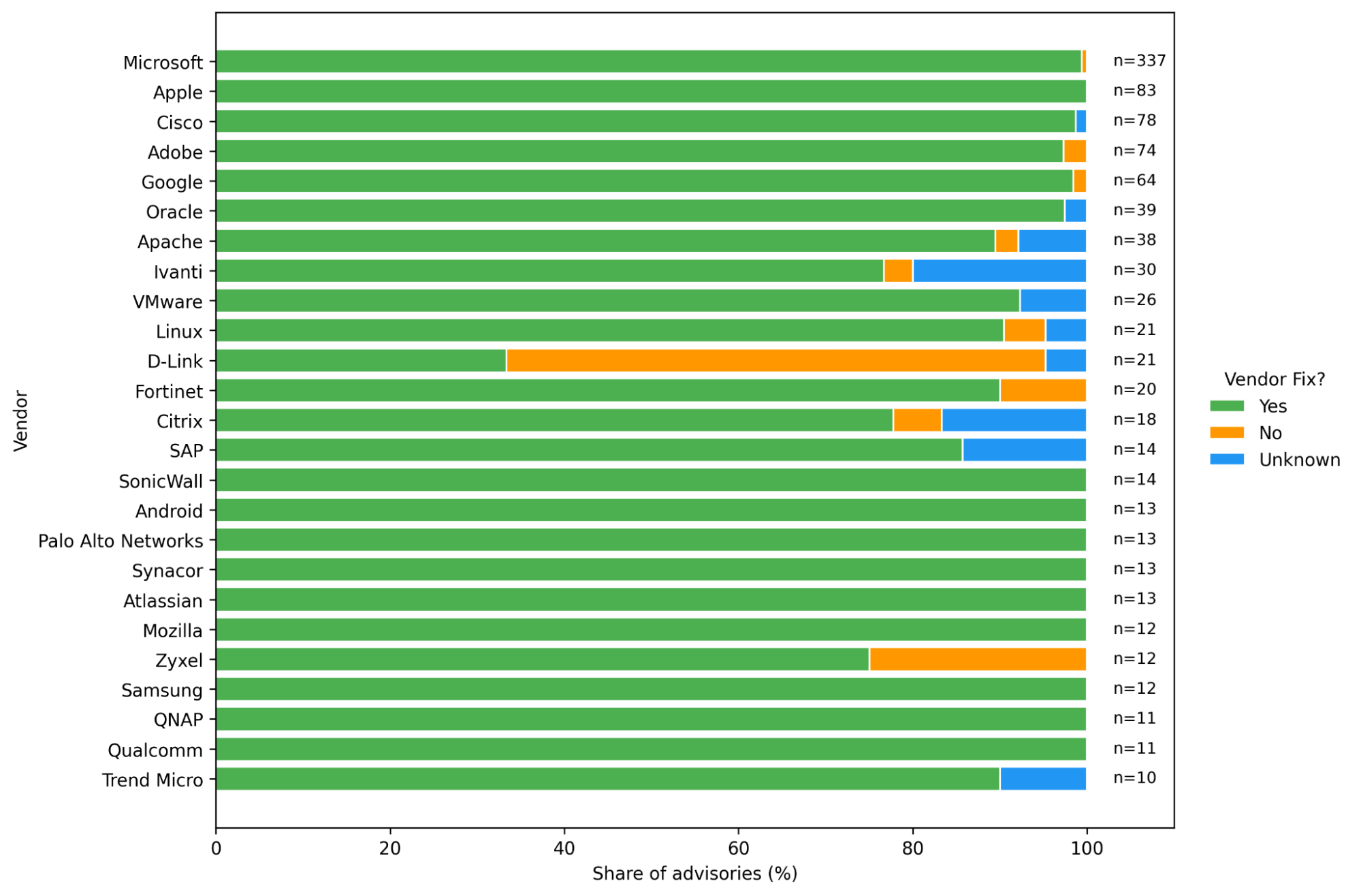}
  \caption{Top vendors in the KEVC and advisory support.}
  \label{fig:vendor-fix}
\end{figure}

We reviewed the KEVC to identify which entries have mapped vendor advisories (Figure~\ref{fig:vendor-fix}). Most vendors do publish advisories, and where none are listed, the cases typically involve end-of-life software (e.g., Adobe Flash), where vendor support is no longer available.


\begin{figure}[htb]
  \centering
  \begin{minipage}{0.48\textwidth}
    \centering
    \includegraphics[width=\linewidth]{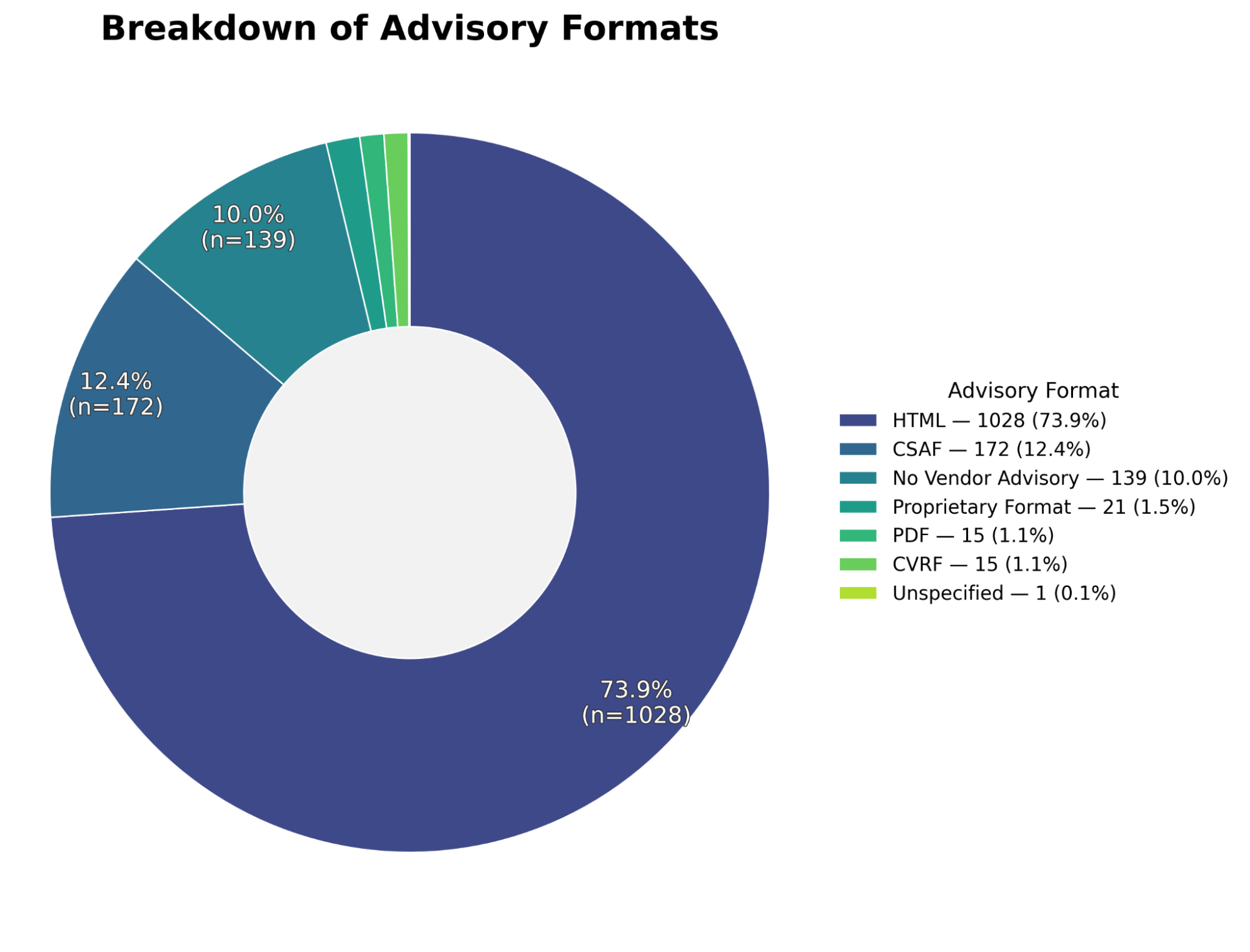}\\
  \end{minipage}\hfill
  \begin{minipage}{0.48\textwidth}
    \centering
    \includegraphics[width=\linewidth]{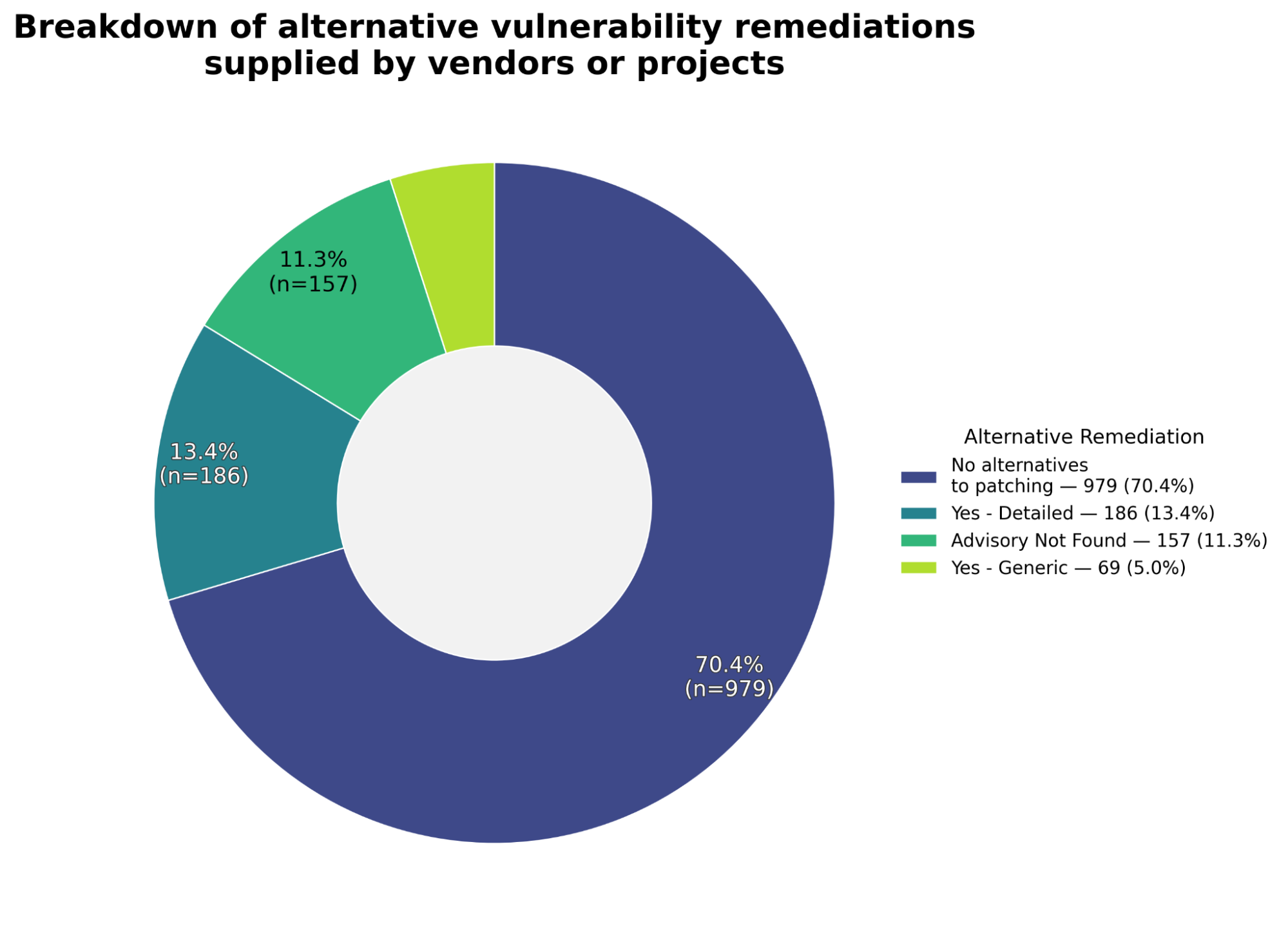}\\
  \end{minipage}
  \caption{Advisory characteristics for KEVC entries. 
(Left) Advisory publication formats associated with KEVC entries. 
(Right) Presence and quality of workarounds and mitigations in advisories.}
  \label{fig:kevc-advisories}
\end{figure}

However, 73.9\% of advisories are provided as web pages rather than in a machine-readable format such as CSAF (12.4\%) or legacy CVRF (1.1\%) (Figure~\ref{fig:kevc-advisories}, left). This restricts automated mapping between advisories and vulnerabilities and makes analysis more manual and error-prone.

Substantively, 81.7\% of vulnerabilities either do not have an associated advisory (11.3\%) or the advisories do not include any workaround or mitigation (70.4\%) (Figure~\ref{fig:kevc-advisories}, right). Among those that do, a small portion (5.0\%) provide guidance that is too generic (e.g., broad ``best practices'' lists) to be implemented as concrete, compensating controls. Only 13.4\% offer specific, actionable mitigations suitable as patch alternatives in OT.

Thus, while advisories exist for most KEVC entries, their format and content often fall short of OT needs. Improving machine-readable publication (e.g., CSAF with structured mitigation fields) and increasing the proportion of vulnerabilities having patching alternatives would materially strengthen remediation pathways in environments where immediate patching is not possible.

\subsection{Augmenting Advisories through Exploit Analysis}
Given that 87\% of vulnerabilities in the KEVC provide no specific remediation guidance, we tested whether exploit analysis could supply OT-appropriate patch alternatives. For an OT-relevant KEVC subset, we parsed vulnerability and exploit descriptions, mapped them to MITRE ATT\&CK techniques, and then to ATT\&CK mitigations. For each vulnerability, we translated the mapped mitigation into actionable, OT-feasible steps (e.g., zoning/segmenting conduits, deny-by-default ACLs, service hardening, application allowlisting) and rated expected effectiveness as \emph{High} (prevents or reliably blocks the exploit), \emph{Medium} (meaningfully reduces the likelihood or impact), or \emph{Low} (limited effect).

\begin{figure}[htbp]
  \centering
  \includegraphics[width=\linewidth]{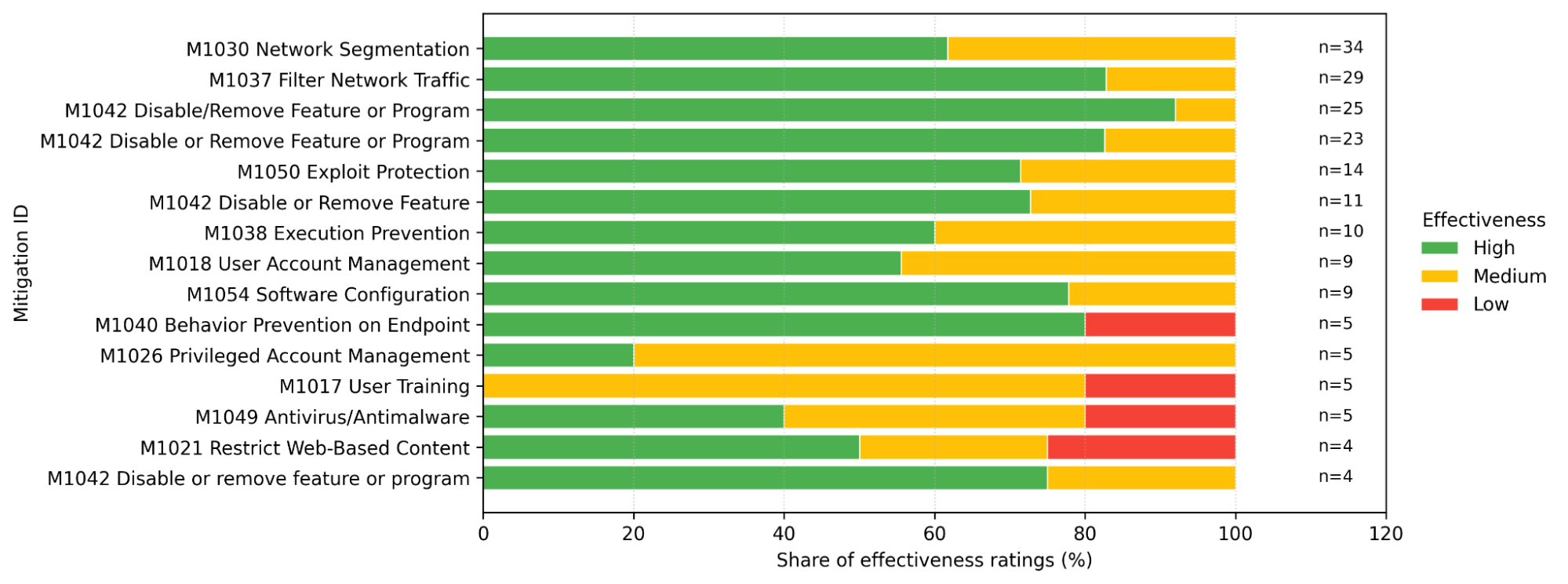}
  \caption{Top patch alternatives derived from exploit analysis mapped to ATT\&CK.}
  \label{fig:custom-mitigation}
\end{figure}

This process yielded an initial 252 tailored workarounds/mitigations covering 198 vulnerabilities. As shown in Figure~\ref{fig:custom-mitigation}, \textit{M1030: Network Segmentation} dominates the top 15 remediations, with over half of the techniques rated \emph{High}. This aligns with our earlier observation that many of the vulnerabilities are non-interactive in nature. In contrast, measures such as \textit{M1017: User Training} and \textit{M1021: Restricting Web-Based Content} tend to score \emph{Low} and \emph{Medium} due to their limited ability to prevent non-interactive network attack paths common in OT.

\section{Implications for OT Cyber Defense}
In OT systems, KEVC triage is a core discipline. Defenders must continuously determine which vulnerabilities are both relevant to the installed base and realistically exploitable under OT operating conditions. Our analysis indicates that KEVC entries frequently touch software present in OT contexts and a nontrivial subset can be exploited with little or no user interaction. This combination raises operational risk while simultaneously narrowing the set of viable responses.

Where vendor advisories omit effective workaround strategies or offer only high-level direction, defenders are not powerless. A practical path is to translate the available exploit descriptions and code into mitigations and workarounds that respect process constraints. Concretely, mapping each CVE to its likely ATT\&CK techniques and then to corresponding ATT\&CK mitigations yields implementable playbook steps that emphasize exposure reduction rather than invasive change. In many cases, zoning and network restrictions, credential and session management, or targeted user awareness training will reduce OT system risk without altering the control application itself. When such measures are recorded with explicit rationale, expected effectiveness, operational change-risk, and validation evidence, they form a defensible catalog that complements vendor guidance.

\section{Future Work}
Advancing this line of work will require coordinated efforts beyond this initial pass. First, the community should formalize and validate two quantitative measures to guide defender decision-making at scale. First, a mitigation effectiveness score will estimate the marginal reduction in exploitation risk from a proposed control leveraging attributes of the vulnerability, mapped attack techniques, and the value density of a successful exploit of the targeted assets. Second, an operational change-risk score should be developed to model the likelihood and consequence of service disruption, including safety impact, downtime requirements, vendor support posture, and test complexity. These measures should be embedded in a machine-readable schema that links KEVC CVEs to concrete remediation playbook steps to enable reproducible results across institutional boundaries.

Entities would also benefit from an open and automated validation toolset that exercises remediation commands and configurations across representative OT platforms (e.g., Windows and Linux engineering workstations, jump hosts, and common network security appliances) in controlled laboratory settings. Public reference implementations and datasets that cover configuration baselines, expected telemetry, and pass/fail criteria would provide a common basis for benchmarking alternative remediations.

Collaboration with vendors and sector information-sharing bodies should prioritize expanding CSAF and VEX fields for workarounds and mitigations so that advisories can be ingested directly into the proposed schema. Standardized fields, examples, and conformance tests could help accelerate ecosystem adoption.

\section{Conclusion}
Focusing on KEVC items is a sound way to contain the attack surface in OT, but it is not sufficient on its own. Advisory formats and content too often stop short of the concrete alternatives that operators need when patching is delayed or impossible. Our results show that meaningful progress is possible by systematically translating exploit behavior into layered operationally feasible controls and documenting those controls as repeatable playbooks. While vendors should expand machine-readable, environment-sensitive workaround and mitigation guidance, it is neither realistic nor necessary to expect them to shoulder the full burden of alternative remediation across diverse industrial environments.

We therefore propose a shared, standards-based model that augments KEVC and vendor advisories with an open, testable mapping from vulnerabilities to ATT\&CK techniques, to mitigations, and finally to concrete playbook actions with each accompanied by objective measures of expected effectiveness and operational change risk. This approach closes a consequential gap between vulnerability awareness and feasible defense to enable OT operators to act quickly and safely when immediate patching is not an option.

\section*{Acknowledgments}
This material is based upon work supported by the U.S.\ Department of Energy under Award Number DE-CR0000031.

We thank the following research analysts for their substantial assistance with data collection and preliminary analysis for this study: John Birmingham, Isaac Ceja, Isaac Gonzales, Matthew Possehl, Johnathan Reese, Victor Vantseyeva, Nishka Gandu, and Carter Wallace.

\bibliographystyle{unsrt}

\end{document}